\documentclass[fleqn,10pt]{wlscirep}

\usepackage{amsmath}
\usepackage{graphicx}\usepackage{epstopdf} 
\usepackage{filecontents}

\title{Localized Single Frequency Lasing States in a Finite Parity-Time Symmetric Resonator Chain }

\author[1,*]{Sendy Phang}
\author[1]{Ana Vukovic}
\author[2]{Stephen C. Creagh}
\author[1]{Phillip D. Sewell}
\author[2]{Gabriele Gradoni}
\author[1]{Trevor M. Benson}
\affil[1]{George Green Institute for Electromagnetics Research, University of Nottingham, Nottingham NG7 2RD, UK}
\affil[2]{School of Mathematical Sciences, University of Nottingham, Nottingham NG7 2RD, UK}

\affil[*]{sendy.phang@nottingham.ac.uk}

\begin{abstract}
In this paper a practical case of a finite periodic Parity Time chain made of resonant dielectric cylinders is considered. The paper analyzes a more general case where PT symmetry is achieved by modulating both the real and imaginary part of the material refractive index along the resonator chain. The band-structure of the finite periodic PT resonator chains is compared to infinite chains in order to understand the complex interdependence of the Bloch phase and the amount of the gain/loss in the system that causes the PT symmetry to break.  The results show that the type of the modulation along the unit cell can significantly affect the position of the threshold point of the PT system. In all cases the lowest threshold is achieved near the end of the Brillouin zone. In the case of finite PT-chains, and for a particular type of modulation, early PT symmetry breaking is observed and shown to be caused by the presence of termination states localized at the edges of the finite chain resulting in localized lasing and dissipative modes at each end of the chain. 
\end{abstract}
\begin{document}

\flushbottom
\maketitle
%
%
\thispagestyle{empty}

\section*{Introduction}

\textbf{This paper focuses on the resonant modes of a finite chain of PT-resonant cavities. In particular, we report the existence of termination states, which are highly localized at the ends of such structures and which are topologically distinct from states obtained from the band structure of the infinite-chain case. These termination states are observed when the refractive index has appropriate periodic modulation and exhibit lasing behavior for an arbitrarily small values of gain/loss. They are therefore very interesting for the development of single-frequency laser cavity applications.}

Since the conceptualization of the parity-time (PT) symmetric Hamiltonian in quantum mechanics\cite{Bender1999}, photonics has been proved to be an excellent platform for exploring practical applications arising from a variety of PT structures. A main feature of PT-symmetric photonic structures is that they may have purely real spectra, i.e. zero net-power amplification or dissipation, despite having both gain and loss in the system. However, there exists a threshold defined for a certain amount of gain/loss for which the PT-system undergoes a spontaneous PT-symmetry breaking, and above which the eigenfrequencies become complex and power grows exponentially \cite{Makris2008,Regensburger2012,chong2011,Ge2012,Ruter2010,Mostafazadeh2012} . Interesting properties of PT-photonic structures include loss-induced invisibility\cite{Lin2011}, simultaneous lasing and coherent perfect absorption\cite{chong2011}, anomalous Bloch-mode power oscillation\cite{Makris2008,Regensburger2013,Longhi2009}, asymmetric beam scattering\cite{Lin2011,chong2011,Ruschhaupt2005,Chang2014,Peng2014d,Feng2013}  and loss-induced lasing\cite{Peng2014,Brandstetter2014,phang2015b}. Based on the PT-symmetric concept, several photonic structures have been studied both theoretically and experimentally in recent years, including gratings\cite{Lin2011,Phang2013,Phang2014d,Phang2015,Phang2014,Rivolta2015,Ramezani2010,longhi2010a,Kulishov2005,Kulishov2013}, lattices\cite{Regensburger2013,Makris2008,Zheng2010,Dmitriev2010,baras2013}, waveguides\cite{Ruschhaupt2005,Ruter2010,Ctyroky2014,Kuzmiak2010,Lupu2013,Greenberg2005,Phang2014c,Nolting1996}, plasmonics\cite{Benisty2011,Baum2015,Alaeian2014} and resonant cavities\cite{Feng2014,Longhi2014,Chang2014,Feng2013,Peng2014,Hodaei2014a,Brandstetter2014,phang2015b}. Applications such as switches\cite{Nazari2011,Lupu2013,Phang2013}, memory devices\cite{Kulishov2013,Phang2014d}, logic gates\cite{Phang2015}, and lasers\cite{Longhi2014,Peng2014,Feng2014} have been proposed to harness the unusual properties of PT structures. 

Although there have been extensive studies of PT-symmetric system based on waveguiding structures\cite{Regensburger2013,Makris2008,Zheng2010,Dmitriev2010,baras2013}, it is just recently that PT-symmetric resonant structures have taken the stage\cite{Feng2014,Longhi2014,Chang2014,Feng2013,Peng2014,Hodaei2014a,Brandstetter2014,phang2015b}. Utilizing the attributes of a resonant structure, operations such as strong field localization, energy build-up and frequency trapping are enabled\cite{Boriskina2006,Smotrova2006}. In contrast to a PT-symmetric waveguide system, in which the eigenmodes are real during the PT-symmetric phase and form a complex conjugate pair during the PT broken-symmetry phase, two-coupled resonant cavities, forming a simple PT resonant-coupler, have been shown to have complex eigenfrequencies even during the PT-symmetric phase, owing to the radiation losses in an open system\cite{Feng2014,Chang2014,Feng2013,Peng2014,Brandstetter2014,phang2015b}. Although the PT resonant-coupler exhibits unbalanced gain and loss, the feature of spontaneous breaking-symmetry remains\cite{Feng2014,Chang2014,Feng2013,Peng2014,Brandstetter2014,phang2015b}.  

Moreover, to date only periodic lattices of PT waveguides structures\cite{Regensburger2013,Makris2008,Zheng2010,Dmitriev2010,baras2013} have been analyzed and shown to exhibit Bloch mode-power oscillation, double refraction\cite{Makris2008} , band-manipulation\cite{Regensburger2013}  and recently a long-lived defect state\cite{Weimann2015}. PT-resonant structures considered to date have been exclusively focused on a system of two coupled resonators and where the PT symmetry was induced by modulating the imaginary part of the resonator refractive index, i.e. each resonator had either loss or gain and the real part of resonator refractive index was the same. In such a structure, loss-induced lasing\cite{Peng2014,phang2015b}, opposite lasing and gain-pump dependence behavior \cite{Brandstetter2014} and, with additional passive input/output port waveguides, an asymmetric response depending on the excitation port, have been demonstrated\cite{Peng2014d,Chang2014}. 

In our recent work on PT-coupled resonator cavities\cite{phang2015b}, we have developed an analytical representation of the coupling mechanism, based upon Green’s boundary integral equations (BIE) for a weakly coupled system, to study the spectral properties of such structures under equal gain/loss and fixed-loss-variable-gain scenario. Moreover we demonstrated an induced PT symmetry-breaking phenomenon by increasing loss in the system, which suggests a counterintuitive concept of laser operation in a system dominated by loss rather than gain. Here, we further extend our BIE method to analyze a finite PT periodic chain structure comprised of two-dimensional resonant cavities each supporting whispering-gallery modes. The PT symmetry is introduced by either having a chain of resonators with constant refractive index and an alternating arrangement of gain/loss resonators, or in a more general case where both real and imaginary parts of the resonator refractive index are modulated. 

\section*{Finite PT-Chain Model and Termination States}
In this section we outline the essential features and notation for the finite-chain models under investigation and summarize one of the key features we find in their solution: that when there is an appropriate modulation between resonators, then among the global modes of the system there are distinctive \textit{termination states}. Unlike most global modes, which can be understood as simply discretely sampling the band-structure of the infinite-chain case, these termination states are localised near the ends of the finite chain and have resonant frequencies lying inside the band-gaps of the infinite case. In this section we restrict our discussion to the main qualitative conclusions while in the remainder of the paper we provide a more detailed discussion of the corresponding solutions.

\begin{figure}
\centering
\includegraphics[width=0.5\linewidth]{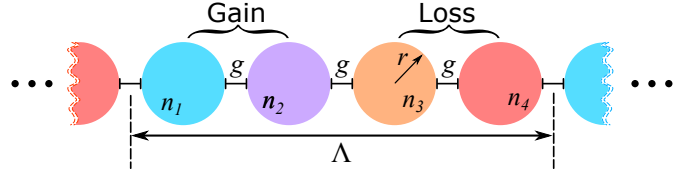}
\caption{\textbf{Schematic illustration of the unit cell of the PT resonant chain.}}
\label{fig:illustrations}
\end{figure}

We begin by describing the model used. In order to satisfy the condition of PT-symmetry\cite{longhi2010b,Longhi2011}, the refractive index profile should satisfy the condition $n(x)=n^* (-x)$, where $*$ denotes complex conjugate. This means that the real part of the refractive index is an even function and the imaginary part of the refractive index is an odd function in space. To capture this condition we consider a unit cell that is comprised of four resonant cavities as shown schematically in Fig. \ref{fig:illustrations}. The unit cell is of length $\Lambda$, all the resonant cavities have the same radius $r$ and are separated by the same gap $g$.
The complex refractive index in each resonant cavity is given by, 
\begin{align}
	\begin{cases}
	n_1 = n_0 +\Delta n'+jn''\\
	n_2 = n_0 -\Delta n'+jn''\\
	n_3 = n_0 -\Delta n'-jn''\\
	n_4 = n_0 +\Delta n'-jn''
	\end{cases}
	,
\end{align}
where $n_0$ is the average real refractive index, $\Delta n'$ denotes the modulation of the real part and $n''$ is the imaginary part of the refractive index such that $n''>0$ represents gain and $n''<0$ represents loss. In this paper, a TM polarized wave is considered where the electric field is directed along the longitudinal resonator axis and it is assumed that the system is weakly coupled i.e. the coupling is considered to be only between nearest neighbors through the evanescent field.

Detailed calculations provided in this paper for the case of finite PT periodic resonator chains comprised of $N=6$ unit cells, each containing four resonators having the same radius $r=0.54\mu m$, average real refractive index $n_0=3.5$ and separation gap $g=0.3\mu m$. The background material is assumed to be air ($n_b=1$). Corresponding isolated resonators are excited at $f_0=336.85$ THz and support a high $Q$-factor whispering-gallery mode of azimuthal order $m=10$, and radial order $n=1$ with $Q$-factor of $1.05\times10^7$. 

\begin{figure} []
\centering
\includegraphics[width=0.8\linewidth]{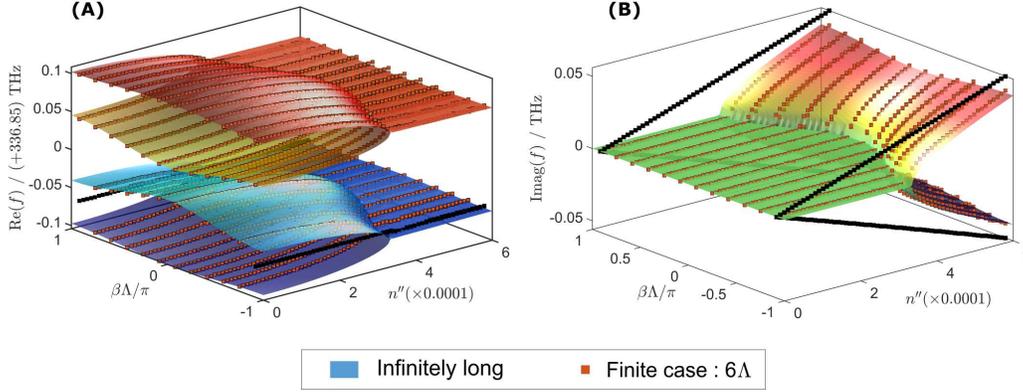}
\caption{\textbf{Band structure diagram and termination states of finite PT-chain.} (A) Real and (B) imaginary part of the eigenfrequencies for the infinite chain (surface plot) and finite PT-chain (discrete square bullets) as a function of Bloch phase $\beta\Lambda$ and gain/loss parameter $n''$ and for $\Delta n'=0.0005$. The termination eigenfrequencies are depicted as the black bullets.}
\label{fig:infiniteandfinite6periodicitywithmodulationp0_0005}
\end{figure}

We now describe the qualitative features of the termination states and state how they are related to the band structure of the infinite-chain limit. Figure \ref{fig:infiniteandfinite6periodicitywithmodulationp0_0005} illustrates how the band-structure of a chain of resonators evolves as the gain/loss parameter $n''$ is increased: the case shown corresponds to coupling of whispering-gallery modes for a sequence of refractive indices given in (1) with real modulation $\Delta n' = 0.0005$.

Real and imaginary parts of the resonant frequency are shown respectively in parts (A) and (B) of Fig. \ref{fig:infiniteandfinite6periodicitywithmodulationp0_0005} as a function of $n''$ and of the Bloch phase $\beta\Lambda$, which is defined so that the solution repeats with a phase delay $\exp(j\beta\Lambda)$ from one unit cell to the next, and where $\Lambda$ is the physical length of a unit cell as illustrated in Fig. \ref{fig:illustrations}. See the appendix for a detailed description of the structure of this state and the method of solution used to find it. In Fig. \ref{fig:infiniteandfinite6periodicitywithmodulationp0_0005}, the band-structure of the limiting infinite-chain limit is shown as a surface. This surface shows threshold behavior: for each value of $\beta\Lambda$, as the gain/loss parameter $n''$ is increased, the real parts of the resonant frequencies approach and coalesce at a critical value (which is dependant on $\beta\Lambda$), after which the imaginary parts split and become significantly complex.  This is commonly referred to as the PT-symmetric threshold, beyond which the structure is in the PT broken-symmetry phase\cite{chong2011,Ge2012}. For the case where the real part of the refractive index is modulated, as it is here, we find that there is a distinct threshold (at nonzero values of $n''$) for all values of $\beta\Lambda$. We will see in the more detailed discussion of the following sections that when the modulation is decreased, this threshold also decreases, so that when there is no modulation at all, the eigenvalues are thresholdless for some values of $\beta\Lambda$. It is important to note that, there are multiple band-gaps formed between the band surfaces when the gain/loss parameter $n''$ is below the threshold as seen in Fig. \ref{fig:infiniteandfinite6periodicitywithmodulationp0_0005}.

\begin{figure} [hb]
	\centering
	\includegraphics[width=0.5\linewidth]{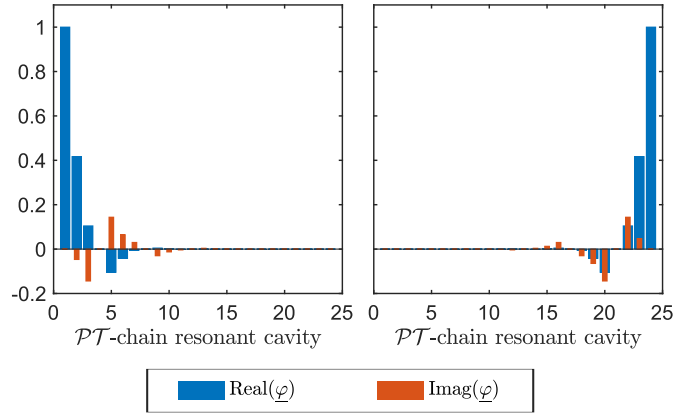}
	\caption{\textbf{Degree of excitation of the termination state within the finite PT-chain.} Eigenvector of the termination states with operated at gain/loss parameter $n''=0.0004$. (Left) Eigenvector for the lasing threshold states ($\text{Im}(f)<0$) and (right) for the dissipative threshold states ($\text{Im}(f)>0$)}
	\label{fig:bar_plot_eigenvector_imag_index_0_0004_with_modulation_dn0_0005}
\end{figure}

Most eigenvalues of the finite PT-chain case can be understood by sampling this band surface at discrete values of the Bloch phase $\beta\Lambda$, as indicated by the bullets on Fig. \ref{fig:infiniteandfinite6periodicitywithmodulationp0_0005}. In addition, however, there are eigenvalues that are completely separate from the band structure of the infinite case. These correspond to the termination states and are denoted using the black bullets in Fig. \ref{fig:infiniteandfinite6periodicitywithmodulationp0_0005}. While typical eigenvectors behave in a pseudo-Bloch way, changing gradually from one unit cell to the next throughout the chain, the termination states are entirely different. Termination states are localized at one of the ends of the finite chain, as illustrated in Fig. \ref{fig:bar_plot_eigenvector_imag_index_0_0004_with_modulation_dn0_0005}. Figure \ref{fig:bar_plot_eigenvector_imag_index_0_0004_with_modulation_dn0_0005} shows typical eigenvectors corresponding to termination states, in which each component describes the degree of excitation of the corresponding whispering gallery mode in an individual resonator. Note that these are related to each other by application of a PT transformation. In this example, the termination state localized on the left is lasing ($\text{Im}(f)<0$) while the termination state localized on the right is dissipating ($\text{Im}(f)>0$). Noting that the resonators terminating the chain on the left have gain whilst the resonators terminating the chain on the right are lossy.

In subsequent sections, we will discuss in detail the spectral behavior of the finite PT-chain of resonators for two different cases of resonator medium refractive index modulation. In the first case, PT behavior is introduced by periodically modulating only the imaginary part of the refractive index whilst the real part stays constant ($\Delta n'=0,n''\ne0$). We will refer to this as the case of simple PT periodicity. In the second case, a more complex form of PT periodicity is also considered when the modulation of both the real part and imaginary part of the refractive index is present ($\Delta n'\ne0,n''\ne0$). We will refer to this as the general case PT periodicity.

\section*{Simple Finite PT-Chain Case}

We start the more detailed analysis by discussing the special case of a finite PT-chain in the absence of a modulation in the real part of the refractive index, i.e. $\Delta n'=0$ in (1). In this case, we found that spectra of finite the PT-chains are obtained as a discrete sampling of the limiting infinite PT-chain within the Brillouin zone. As such, the structure can be both in a PT-symmetric phase and a symmetry-broken phase depending on the Bloch phase and the amount of gain/loss in the system. It is also important to comment that in the absence of real refractive index modulation, we do not observe the termination states that occur in the general PT-chain case when both real and imaginary parts of the refractive index have modulation.

\begin{figure}[tb]
\centering
\includegraphics[width=0.8\linewidth]{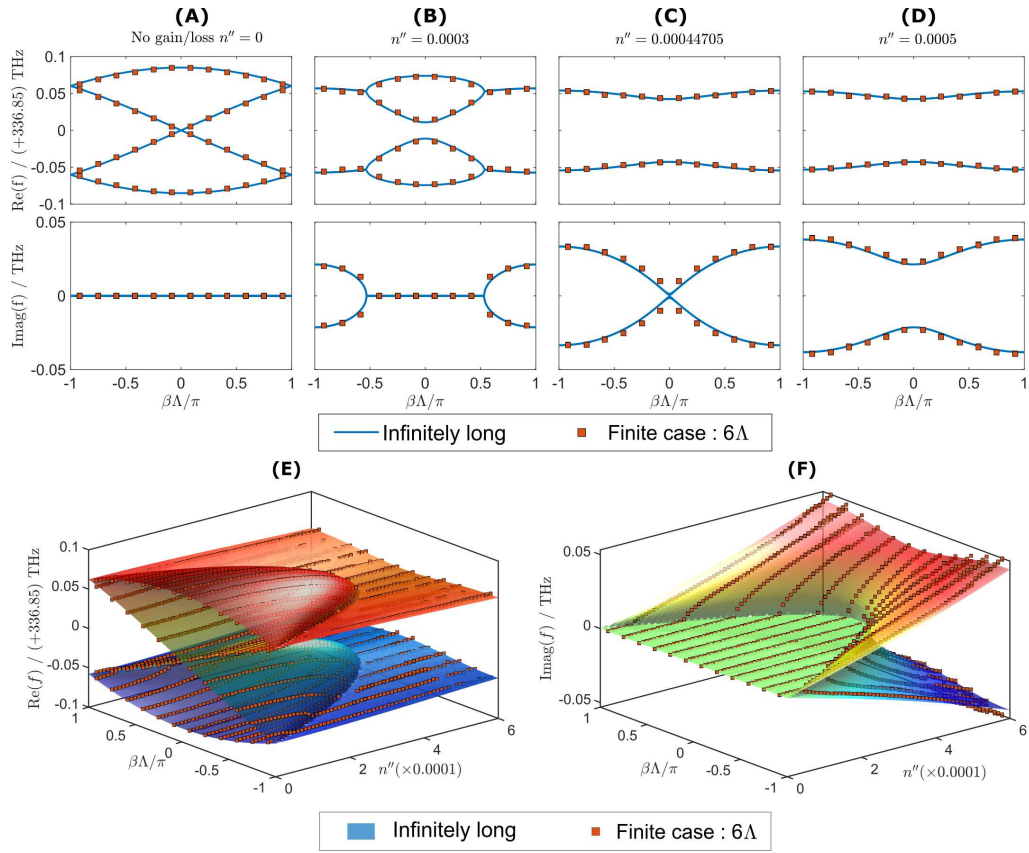}
\caption{\textbf{Band structure diagram of the simple PT-chain resonators.} (A-D) The real part (top panel) and the imaginary part (bottom panel) of the eigenfrequencies obtained for the case of $\Delta n'=0$ for infinite PT chain (solid line) and finite PT chain (discrete square bullets) for: (A) passive chain $n''=0$; (B) $n''=0.0003$; (C) $n''=0.00044705$ and (D) $n''=0.0005$. Part (E) and (F) show the surface plot of the real and imaginary part of the eigenfrequencies as a function of Bloch phase $\beta\Lambda$ and gain/loss parameter $n''$.}
\label{fig:infinite_and_finite_6_periodicity}
\end{figure}

Figure \ref{fig:infinite_and_finite_6_periodicity}(A-D) compares the real and the imaginary parts of the eigenfrequencies of the PT resonator chain for both an infinitely long chain (solid line) and a finite chain consisting of 6 unit cells (discrete points) as a function of the Bloch phase $\beta\Lambda$. Real and imaginary parts of eigenfrequencies are shown in the top and bottom panels respectively. Results are given for different values of gain/loss in the system i.e. $n''=0$, 0.0003, 0.0004472 and 0.0005 respectively. It is emphasized that in the case of no gain/loss, the system does not strictly represent a PT structure as it is an array of identical passive dielectric resonators but is included here for completeness. Figure \ref{fig:infinite_and_finite_6_periodicity}(A) shows that in the case of the passive resonator chain, the real part of the eigenfrequency forms two clusters of modes centered at the operating frequency $f_0$ and the band-structure is symmetrical with respect to the  $\beta\Lambda=0$ axis. In the absence of gain/loss, where $n''=0$, the imaginary part of the eigenfrequency is very small due the coupling between the underlying high $Q$-factor resonator modes (Fig. \ref{fig:infinite_and_finite_6_periodicity}(A) bottom panel), i.e. radiation loss is very small. The real part of the eigenfrequencies of the infinite passive resonator chain shows the presence of degenerate modes at the center,  $\beta\Lambda=0$, and the end of Brillouin zone,  $\beta\Lambda=\pm\pi$.  

In the case of the finite chain with $N=6$ unit cells, (Fig. \ref{fig:infinite_and_finite_6_periodicity}(A)) the eigenfrequencies are discrete. They may be understood as sampling the continuous band-structure at discrete values of $\beta\Lambda=\pm \left(i-\frac{1}{2}\right)\frac{\pi}{N}$ around $\beta\Lambda=0$, where $i=1,2,\cdots,N$. It is worth noting that although these discrete eigenfrequencies follow the general pattern of the eigenfrequencies of the infinite chain, they are not identical. Furthermore, the eigenfrequencies of the finite chain only approach Bloch phases at the points $\beta\Lambda=0$ and $\pm \pi$, as $N$ is increased. 

When the amount of gain/loss in the system is increased to $n''=0.0003$ the band structure is modified in such a manner that the degenerate mode at $\beta\Lambda=0$ splits and forms a bandgap around the frequency $f_0$, as shown in Fig. \ref{fig:infinite_and_finite_6_periodicity}(B). At the same time at the end of Brillouin zone $\beta\Lambda=\pm\pi$, the real values of the eigenfrequencies coalesce but the imaginary parts split and form complex conjugate pair. This shows that the threshold point is determined both by the amount of gain/loss in the system and the Bloch phase. The PT-symmetric region corresponds to where the eigenfrequencies are approximately real. The PT-broken-symmetric phase corresponds to where the eigenfrequencies become complex-conjugate pairs. This is in agreement with the case of PT periodic waveguide lattices, except that in the case of PT periodic resonator chains the eigenfrequencies have a lossy offset\cite{Feng2014,Chang2014,Feng2013,Peng2014,Brandstetter2014,phang2015b}. This is due to the fact that a resonant structure is inherently radiative.

A further increase of gain/loss causes more modes to be in the PT broken-symmetry phase. The eigenfrequencies for a critical value of gain/loss $n''=0.0004472$ are presented in Fig. \ref{fig:infinite_and_finite_6_periodicity}(C). In this specific case, the top panel shows that all the real parts of the eigenfrequencies have coalesced whilst the imaginary parts are split everywhere except at $\beta\Lambda=0$. It is important to note that increase of gain/loss beyond $n''>0.0004472$, causes all eigenfrequencies to occur in complex conjugate pairs and hence the system is in a completely PT-broken symmetry phase, as depicted in Fig. \ref{fig:infinite_and_finite_6_periodicity}(D) for $n''=0.0005$. 

As above, Figs. \ref{fig:infinite_and_finite_6_periodicity}(E) and \ref{fig:infinite_and_finite_6_periodicity}(F), show the band surface plots of the real and imaginary part of eigenfrequencies as a function of both gain/loss parameter $n''$ and the Bloch phase $\beta\Lambda$. The eigenfrequencies of the infinite PT chain are plotted as a surface plot while the eigenfrequencies of the finite PT chain are plotted as discrete bullet points. From Fig. \ref{fig:infinite_and_finite_6_periodicity}(E), it can be seen that for the infinite-chain case, the degenerate mode at $\beta\Lambda=0$ instantaneously splits to form a forbidden-gap around $f_0$ whilst the high and low frequency eigenfrequencies bands coalesce starting from the edge of Brillouin zone $\beta\Lambda=\pm \pi $ towards the center of the Brillouin zone $\beta\Lambda=0$ as the gain/loss is increased. Equally Fig. \ref{fig:infinite_and_finite_6_periodicity}(F) shows that the imaginary part starts splitting from $\beta\Lambda=\pm \pi $ towards $\beta\Lambda=0$ as the gain loss increases, clearly showing that even a small amount of gain/loss can cause PT symmetry breaking in this case.  


Moreover Fig. \ref{fig:infinite_and_finite_6_periodicity}(E) and \ref{fig:infinite_and_finite_6_periodicity}(F) show that the infinite PT chain is a thresholdless lasing structure when operated at the end of Brillouin zone ($\beta\Lambda=\pm \pi $). However, a practical finite PT chain will require small amount of gain/loss to cause PT-symmetry breaking at $\beta\Lambda=\pm \left(\frac{2N-1}{2N}\right)\pi$.  We refer to this \textit{minimum threshold} gain/loss in the finite chain as $n''_{th}$ which is discussed further in the next section.

\section*{Finite PT-Chains with Real Modulation}
This section focuses on the general finite PT-chain when modulation of both real and imaginary part of the refractive index are present, i.e. ($\Delta n'\ne 0$ and $n''\ne 0$) in (1). We find that, by introducing real refractive index modulation, field localization is achievable which leads to a formation of termination states. Furthermore, a judicious distribution of gain/loss causes the termination states to be in the PT broken-symmetry phase, which localizes the solution of the lasing and dissipative termination states at either end of the finite PT-chain structure. 

\begin{figure}
\centering
\includegraphics[width=0.9\linewidth]{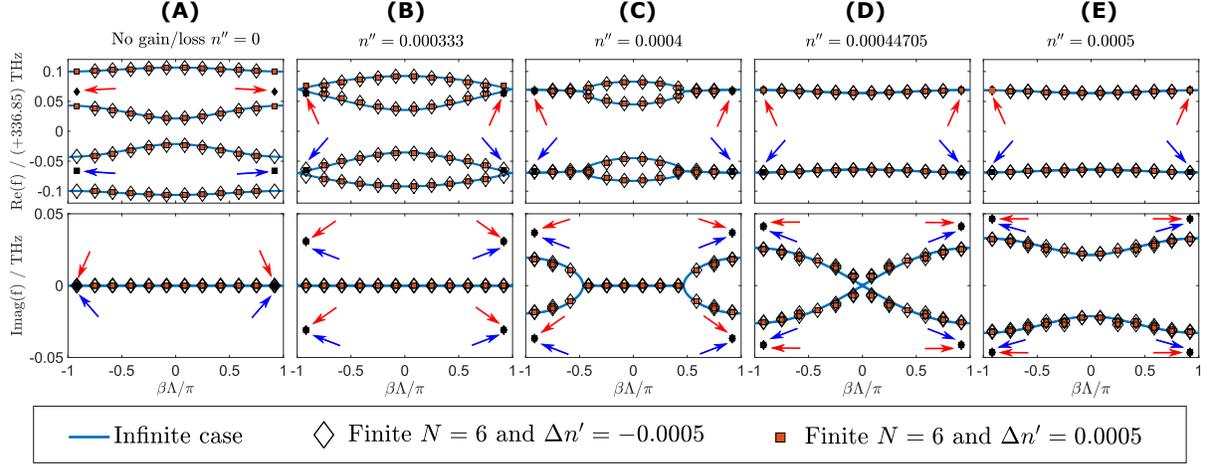}
\caption{\textbf{Band structure diagram of finite PT-chain with modulation at different gain/loss parameter.} The real part (top panel) and the imaginary part (bottom panel) of the eigenfrequencies for the infinite (solid line) and finite (discrete points) PT chain as a function of Bloch phase $\beta\Lambda$ for different values of gain/loss $n''$: (A) passive chain $n''=0$; (B) $n''=0.000333$; (C) $n''=0.0004$, (D) $n''=0.00044705$ and (E) $n''=0.0005$. Real refractive index modulation $\Delta n'=-0.0005$ and 0.0005 are represented by diamond and square points respectively. The blue and red arrows points to the termination states of the negative and positive modulation respectively.}
\label{fig:infinite_and_finite_6_periodicity_with_modulation_slices}
\end{figure}

We will consider a unit cell of the PT chain which is modulated in the manner of (1) with $\Delta n'=\pm 0.0005$, so that it satisfies the PT condition, i.e. that the real part of the refractive index profile is an even function, while the imaginary part of the refractive index is an odd function in space. Figure \ref{fig:infinite_and_finite_6_periodicity_with_modulation_slices} compares the corresponding band-structure of the infinite (solid line) and finite PT-chains with $N=6$ unit cells (discrete bullet points) for different values of the gain/loss parameter $n''$. It is emphasized here that the sign of the real index modulation will not affect the eigenfrequencies of the infinite chain as it results in identical set of coupled equations (2). However, for the finite PT chain case, the sign of modulation \textit{will} affect the position of the eigenfrequencies which are denoted as diamonds for $\Delta n'=-0.0005$ and squares for $\Delta n'=+0.0005$. 

Figure \ref{fig:infinite_and_finite_6_periodicity_with_modulation_slices}(A) shows the eigenfrequencies of the infinite and finite long passive resonator chain, i.e. $n''=0$. The band dispersion structure shows three band-gaps in the real eigenfrequenices whilst the imaginary part of eigenfrequencies are almost zero for all values of the Bloch phase.  Comparing this results with Fig. \ref{fig:infinite_and_finite_6_periodicity}(A), it can be seen that introduction of the real index modulation has caused the splitting of degenerate eigenfrequencies at $\beta\Lambda = 0$ and $\pm \pi $ thus creating three forbidden band-gaps where there is only one in Fig. \ref{fig:infinite_and_finite_6_periodicity}(A). In the case where the finite resonator chain is passive, the discrete eigenfrequencies mainly follow the path of the eigenfrequencies of the infinite resonator chain, with the exception of the degenerate modes at $\beta\Lambda = \pm 5.5\pi/6$, which we refer to as the termination states (Fig. \ref{fig:infinite_and_finite_6_periodicity_with_modulation_slices}(A)). In the case of positive real refractive index modulation ($\Delta n'=0.0005$) and for the real values of eigenfrequencies, the termination states occur at the low frequency band cluster and are highlighted by the blue arrows, whilst for the case of negative real refractive index modulation ($\Delta n'=-0.0005$) the termination states occur at the high frequency band-cluster and are highlighted by red arrows. Furthermore, for the finite passive resonator chain, the imaginary parts of the eigenfrequencies are almost zero for both cases of positive and negative real part modulation and are shown by overlapping discrete points in the bottom panel of Fig. \ref{fig:infinite_and_finite_6_periodicity_with_modulation_slices}(A). The imaginary part of the termination states is the same for both positive and the negative modulation of the real part of the refractive index as indicated by the red and blue arrows. 

\begin{figure}[]
\centering
\includegraphics[width=0.5\linewidth]{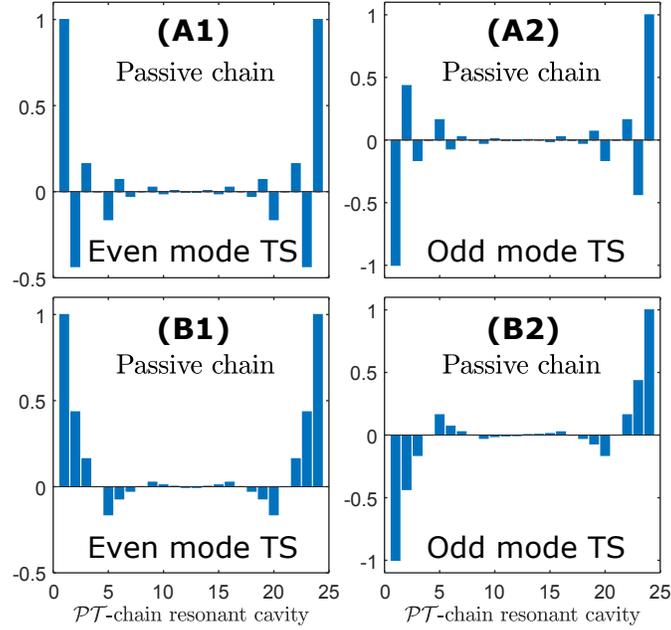}
\caption{\textbf{Distribution of eigenvectors of termination states for the finite passive resonator chain.} The top (A) and bottom (B) panel are for positive real index modulation respectively. The real refractive index modulation considered $\Delta n'=\pm0.0005$. The eigenvector of index (1) associated with the even mode and index (2) associated for the odd termination states (TS).}
\label{fig:bar_plot_eigenvector_with_modulation_dn_pm0_0005_no_gain_loss}
\end{figure}

The eigenvectors associated with the termination states of the finite passive chain of resonators are depicted in Fig. \ref{fig:bar_plot_eigenvector_with_modulation_dn_pm0_0005_no_gain_loss}. As a passive chain, the termination states are a degenerate mode pair which led to the formation of an even and odd spatial termination states. Noting that, the eigenvectors here are associated with the degree of excitation of the whispering-gallery mode distributed within the chain. The eigenvectors show that the termination states are highly localized; the field near the termination is the strongest in amplitude and the field amplitude decreases towards the middle of the chain. Figure \ref{fig:bar_plot_eigenvector_with_modulation_dn_pm0_0005_no_gain_loss}(A) and \ref{fig:bar_plot_eigenvector_with_modulation_dn_pm0_0005_no_gain_loss}(B) depict the eigenvectors for negative and positive real index modulation, respectively, and demonstrate the difference between the two modulations, which cause the resonator to be excited with different phase but with equal strength.     

The band diagram for infinite and finite PT-chain resonators with the gain/loss value $n''=0.000333$ is depicted in Fig. \ref{fig:infinite_and_finite_6_periodicity_with_modulation_slices}(B). The real part of the eigenfrequency is given by the top panel and demonstrates that the forbidden band-gaps at the high and low frequency clusters are reduced as the real part of eigenfrequencies coalesce at $\beta\Lambda = \pi$. Most of the eigenfrequencies of the finite PT chain mimic the behavior of the eigenfrequencies of the infinite PT chain with the exception of the eigenfrequencies of the termination states. The real parts of the eigenfrequencies of termination modes coalesce whilst the imaginary parts split into complex conjugate pairs as shown in Fig. \ref{fig:infinite_and_finite_6_periodicity_with_modulation_slices}(B) for $n''=0.000333$. It is important to note that although the values of real parts of the eigenfrequency of the termination states are approximately equal, they are not degenerate because the imaginary parts of the eigenfrequency are different.   


With a further increase of gain/loss $n''$, in the case of both infinite and finite PT resonator chains, more states coalesce in both the high and low frequency band-clusters from $\beta\Lambda = \pm \pi$ towards $\beta\Lambda = 0$. Correspondingly, the imaginary part splits to form pairs of complex conjugates eigenfrequencies, as shown in Fig. \ref{fig:infinite_and_finite_6_periodicity_with_modulation_slices}(C,D). Operation with gain/loss $n''>0.00044705$ leads to an operation with completely complex conjugated eigenfrequencies, indicating that the system is completely in the PT broken-symmetry phase, as shown in Fig. \ref{fig:infinite_and_finite_6_periodicity_with_modulation_slices}(E). Finally, it can be observed from Fig. \ref{fig:infinite_and_finite_6_periodicity_with_modulation_slices}(B)-(E) that the imaginary parts of the termination states continue to increase in value as gain/loss increases in the system.  

The eigenvectors of the termination states of the finite PT-chain are depicted in Fig. \ref{fig:bar_plot_eigenvector_imag_index_0_0004_with_modulation_dn0_0005} for the positive sign of real index modulation and for gain/loss value $n''=0.0004$. Comparing the eigenvectors of the termination states for the different signs of real index modulation, it is observed that the amplitudes of the eigenvectors are the same but have different phase. This result, which is not shown separately in this paper, confirms that for both cases the lasing termination states are localized at the left end of the chain whilst the dissipative termination states are localized at the right end. It is important to note that the eigenvectors of the lasing and dissipative termination states are related by the PT transformation.

In practice, these termination states manifest themselves as localized lasing or dissipating modes at either termination end of the PT chain. The imaginary part of the eigenvectors starts to increase as the gain/loss parameter $n''$ increases. It is worth commenting that these termination states are similar to the topological states induced in the PT two-dimensional honeycomb photonic crystal lattices system\cite{Harari2015a} which are immune to the presence of a defect. Both termination and topological states exist on the edges of the periodic medium, but while topological states propagate around the edge of the structure, the termination modes in our configuration are stationary and have either lasing or dissipating behavior at different ends of the chain.


\begin{figure} []
	\centering
	\includegraphics[width=0.8\linewidth]{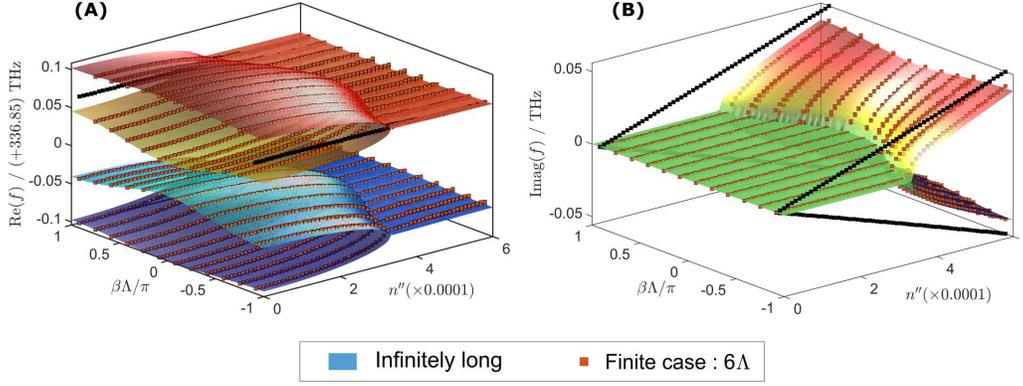}
	\caption{\textbf{Band structure diagram and termination states of finite PT-chain with negative value real modulation.} (A) Real and (B) imaginary part of the eigenfrequencies for the infinite (surface plot) and finite PT-chain (discrete square bullets) as a function of Bloch phase $\beta\Lambda$ and gain/loss parameter $n''$ and for $\Delta n'=-0.0005$. The termination states are depicted as the black bullets. }
	\label{fig:infiniteandfinite6periodicitywithmodulationn0_0005}
\end{figure}

Figure \ref{fig:infiniteandfinite6periodicitywithmodulationn0_0005} shows the surface plots of the real and imaginary parts of eigenfrequencies of the PT chain, with a real refractive index modulation of $\Delta n'=-0.0005$, as a function of both gain/loss parameter $n''$ and the Bloch phase $\beta\Lambda$. These results differ from the case illustrated in Fig. \ref{fig:infiniteandfinite6periodicitywithmodulationp0_0005}(A,B) by having a negative rather than a positive value of $\Delta n'$. Again, the eigenfrequencies of the infinite PT chains are plotted as a surface plot while the eigenfrequencies of the finite PT chain are plotted as discrete points. It is noticeable that the infinite PT-chain with a general modulation has a significantly different band-structure compared to the band-structure for the case of simple PT modulation shown in Fig. \ref{fig:infinite_and_finite_6_periodicity}(E,F). Prominent forbidden band-gaps appear in the PT-chain with real index modulation; these band-gaps cause symmetry-breaking to occur at much higher values of gain/loss $n''$. However, the main difference is the presence of the termination states which are marked by black bullets in all plots. It can be seen that for the case of negative modulation the termination states occur at the high frequency band-cluster, as shown in Fig. \ref{fig:infiniteandfinite6periodicitywithmodulationn0_0005}(A), while for the positive modulation the termination eigenfrequencies are located at the low frequency band-cluster as shown in Fig. \ref{fig:infiniteandfinite6periodicitywithmodulationp0_0005}(A). It can also be observed that for the Bloch phase $\beta\Lambda=\pm 5.5\pi/6$, the termination states undergo spontaneous PT symmetry-breaking. This is almost thresholdless as the imaginary part splits into pairs of complex conjugate eigenfrequencies immediately once gain/loss is present, see Fig. \ref{fig:infiniteandfinite6periodicitywithmodulationp0_0005}(B) and \ref{fig:infiniteandfinite6periodicitywithmodulationn0_0005}(B). This almost zero gain/loss threshold indicates that a finite PT-chain can support lasing modes with very low amounts of added gain/loss in the system. The lasing mode is highly localized at one end of the chain. In order to reach the next gain/loss threshold much higher gain/loss is needed at which point the system reaches the PT symmetry-broken phase when more modes start to lase. 


\begin{figure}[h]
	\centering
	\includegraphics[width=0.7\linewidth]{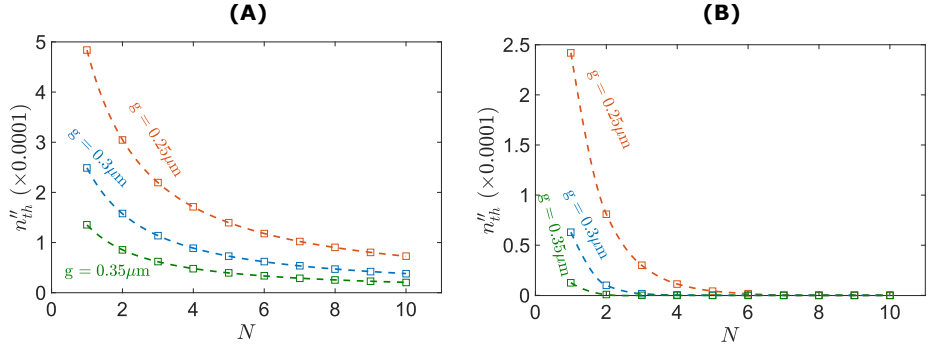}
	\caption{\textbf{Threshold point of the termination states of the finite PT chain.} The impact of the number of unit cell $N$ and separation gap $g$ to the critical gain/loss parameter $n''_{th}$. For (A) the simple PT-chain case and (B) for the general PT-chain case with real modulation $\Delta n'=0.0005$}
	\label{fig:initial_treshold_different_gap_different_no_periods}
\end{figure}

Closer investigation of the threshold points of the finite PT-chains in Fig. \ref{fig:infinite_and_finite_6_periodicity} and Fig. \ref{fig:infinite_and_finite_6_periodicity_with_modulation_slices} suggests that the termination states undergo a PT symmetry-breaking at lower values of gain/loss than for the case of the PT-chain with simple modulation.  To confirm this observation, the minimum threshold $n''_{th}$ needed to break the PT-symmetry of the termination states of the simple and PT-chain with real modulation are plotted in Fig. \ref{fig:initial_treshold_different_gap_different_no_periods}(A) and (B), respectively, as the function of number of unit cells and the separation gap $g$. As the PT symmetry-breaking occurs simultaneously for both positive and negative real index modulation in the general PT-chain case, we only consider case of $\Delta n'=0.0005$. By comparing Figs. \ref{fig:initial_treshold_different_gap_different_no_periods}(A) and \ref{fig:initial_treshold_different_gap_different_no_periods}(B), it can be observed that the gain/loss needed to cause PT symmetry-breaking is lower in the case of PT-chain with real modulation. Furthermore, it can also be seen that the rate at which minimum threshold decreases is faster for the PT-chain with real modulation. The minimum threshold can be further reduced by increasing the separation gap $g$ between the resonators, which decreases the coupling strength between the resonators.

\section*{Conclusion}
This paper analyzes the spectral properties of finite PT-symmetric chains of dielectric resonators. The PT-symmetry is introduced either as a simple modulation of resonators’ refractive index along the chain i.e., as a chain of alternating gain and loss resonators or, by additionally introducing a modulation of the real part of the resonator refractive index so that PT symmetry condition is satisfied. In order to consider the more general PT symmetry condition a unit cell of four dielectric resonators is considered. The results show that in the case of simple PT modulation, the infinite PT resonator chain has a zero PT threshold whilst the finite PT resonator chain needs a certain critical gain/loss to achieve PT-symmetry breaking. Furthermore the band-structure of the finite PT-chain is a discretely sampled limit of the infinite PT-chain case. 

In the case of general PT symmetry the band-structure shows clear band-gaps for both infinite and finite PT-chains. Furthermore the finite PT-chain shows the existence of termination states, which have their highest field intensity localized near the termination ends of the resonator chain. Although general PT symmetry-breaking occurs at much higher values of gain/loss compared to PT-chains with simple modulation, the presence of the termination states in the practical finite-chain causes an almost thresholdless PT symmetry-breaking. This PT-breaking is now localized to the edge resonators, with one end of the chain is lasing and the other dissipating. Significantly higher gain/loss is needed to achieve complete PT-symmetry breaking in the case of PT resonator chains with general PT symmetry, indicating that the region in between localized PT-symmetry can be utilized for lasing applications. 

In both cases the amount of the threshold gain can be further reduced by reducing the coupling in the system, for example by reducing the resonator separation or by increasing the number of unit cells in the finite PT chains.

\section*{Appendix - Analytical Representation of the Coupling System}

In the case of an infinite chain there exists a continuity of eigenstates $\varphi_i$ evolving between the left and right end of the unit cell with a phase delay given by the Bloch phase $\beta \Lambda$\cite{Collin1991}. Using the Bloch theorem, the BIE reduces to $4\times4$ linearly independent equations of the form,
\begin{align}
\begin{pmatrix}
\tilde{D}_1                & \tilde{C}   &   0         & e^{j\beta\Lambda}\tilde{C} \\
\tilde{C}                  & \tilde{D}_2 &   \tilde{C} & 0   \\
0                          & \tilde{C}   & \tilde{D}_3 &  \tilde{C}    \\
e^{j\beta\Lambda}\tilde{C} &   0         & \tilde{C}   & \tilde{D}_4    \\
\end{pmatrix} 
\begin{pmatrix}
\varphi_1 \\
\varphi_2  \\
\varphi_3  \\
\varphi_4
\end{pmatrix} = 0
\end{align}
where, elements $\tilde{D}_{i=1,2,3,4}$ describe the field of an individual isolated resonator and are given by\cite{phang2015b},
\begin{align}
\tilde{D}_{i=1,2,3,4} = 
-j \frac{zH_m(u)J'_m(z_i) - uH'_m(u)J'_m(z_i)}{zJ_m(u)J'_m(z_i) - uJ'_m(u)J_m(z_i)}
\end{align} 
with $z_i=n_i k_0 r$  and $u=k_0 r$, $k_0$ is the free space wavenumber, and $m$ denotes an azimuthal order of the whispering gallery mode.  The elements $\tilde{C}$ represent the field outside the resonator which couples the solution of the neighboring resonator for a given mode $m$ and are 
\begin{align}
\tilde{C} -jH_{2m}(w)
\end{align}
where $w=k_0 (2r+g)$. In (3) and (4), $J_m$ and $J_m'$ denote the Bessel function of order $m$ and its derivative respectively, $H_m$ and $H_m'$ denote the Hankel function of the second kind of order $m$ and its derivative respectively. The eigenfrequencies  are obtained as solutions of eq.(2) which for the PT infinite chain comprised of four resonators results in four eigenfrequencies $f_{i=1,2,3,4}$ associated with the specified whispering-gallery mode $m$ and at desired Bloch phase $\beta\Lambda$.

In the case of the finite chain consisting of $N$ unit cells, the matrix is of order $4N\times4N$ and has the form,
\begin{align}
\begin{pmatrix}
\tilde{D}_1  & \tilde{C}   &   0         & 0           & 0 & 0 \\
\tilde{C}    & \tilde{D}_2 & \tilde{C}   & 0           & 0 & 0 \\
0            & \tilde{C}   & \tilde{D}_3 &  \tilde{C}  & 0 & 0    \\
0            &   0         & \tilde{C}   & \tilde{D}_4 & \tilde{C} & 0  \\
0            &   0         &   0         & \tilde{C}   & \tilde{D}_1 & \ddots   \\
0            &   0         &   0         &   0       & \ddots   & \ddots   
\end{pmatrix} 
\begin{pmatrix}
\varphi_1 \\
\varphi_2  \\
\varphi_3  \\ 
\varphi_4  \\
\varphi_1 \\
\vdots
\end{pmatrix} = 0
\end{align}
The eigenfrequencies $f_i$ for $i=1,2,…,4N$ are obtained by numerically solving (5) for the specified mode number $m$. 

\bibliography{refs}


\section*{Author contributions statement}
S.P. conceived the idea, conducted the calculation and produced the results, S.C.C and S.P developed the model, S.P., A.V. and S.C.C. wrote the manuscript, S.P., A.V., S.C.C, T.M.B, G.G and P.D.S. analyzed the results.  A.V., S.C.C., P.D.S. and T.M.B. supervised the project. All authors reviewed and approved the manuscript. 

\section*{Additional Information}
\begin{description}
	\item[Competing financial interests:] The authors declare no competing financial interests.
\end{description}

\end{document}